\documentclass[preprint,aps]{revtex4}
\usepackage{graphicx}
\usepackage{dcolumn}
\usepackage{bm}

\begin{document}

\title{Comment on ``Far-field microscopy with a
nanometer-scale resolution based on the in-plane image
magnification by surface plasmon polaritons''}

\author{Aur\'{e}lien~Drezet}
\email{aurelien.drezet@uni-graz.at}
\author{Andreas Hohenau}
\author{Joachim R.~Krenn}
\affiliation{Institute of Physics, Karl-Franzens University,
Universit\"atsplatz 5, A-8010 Graz, Austria}
\date{\today}

\maketitle

\maketitle

Recently, Smolyaninov \emph{et al.} [1] reported the realization of
a 2D sub-wavelength resolution microscope using surface plasmon
polaritons (SPP) propagating on a gold film bounded by the glass
substratum and a glycerin droplet [1-3]. The main idea of [1] is
that SPPs should, for a laser wavelength $\lambda_{0}$ of 500 nm,
have a very short wavelength of $\lambda_{\rm SPP}\simeq$ 70 nm
which implies an ultimate resolution of $\lambda_{\rm SPP}/2\simeq
35$ nm. We disagree on the conclusions of [1] because the SPP
wavelength is too large and their propagation length too small to
enable the described in-plane imaging process. Indeed, the claims in
[1] are based on the assumption that the SPP modes confined on a
gold metal film are described by an effective Drude model (see
Fig.~1 of [1,2]) requiring precise compensation of the imaginary
parts in the denominator of Eq.~1 of [1] (Eq.~1 supposes
$\lambda_{\rm SPP}\ll \lambda_{0}$ and
$\epsilon_{gold}\simeq-\epsilon_{d}$). However their analysis here
is questionable since in a real metal (see [4]) losses modify
fundamentally the ideal SPP dispersion relations considered in [1].
In order to show that we calculated numerically the dispersion
relation for SPPs confined on a 50 nm thick film as used in [1,2,3].
Fig.~1 clearly reveals two bound SPP modes which are highly damped
in the optical region $\lambda_0\simeq 500$ nm. As visible in
Fig.~1a the SPP wavelength ($=2\pi/\textrm{Re}\{K_{\rm SPP}\}$) at
such $\lambda_0$ takes values centered at 250 nm to 300 nm while the
propagation lengths ($L_{\rm SPP}=1/(2\textrm{Im}\{K_{\rm SPP}\})$)
reach values between 40 and 250 nm (see Fig.~1b). These values agree
with similar theoretical [5,6] and experimental [7] analysis but
disagree completely with the interpretation presented in [1].
Therefore the explanation of the image formation in Fig. 2 of [1]
cannot be considered as correct. Furthermore one can easily see that
with such a wavelength and propagation length SPPs can not be used
to resolve nanoholes separated by 40 nm as claimed in [1] (see
Figs.~3,4 of [1]). Reducing the film thickness could not solve the
problem. Indeed, $\lambda_{\rm SPP}$ and $L_{\rm SPP}$ (of one given
SPP mode) decrease in parallel monotonously (e.~g., $\lambda_{\rm
SPP}=150$ nm and $L_{\rm SPP}=20$ nm for 20 nm thick films) [6].
Such SPPs are clearly non-propagative ($L_{\rm SPP}<\lambda_{\rm
SPP}$) and primarily represent a channel for ohmic losses and
scattering. There is no hope in such conditions to justify high
resolution by an hypothetical ``mode coupling'' mechanism [1,2]
called to transfer the local information from SPPs to guided waves
in the glycerin as $L_{\rm SPP}$ is simply too small. The additional
high-resolution imaging mechanisms based on the properties of SPP
Bloch waves with short wavelengths suggested in [2, 8] do not appear
more convincing since i) it can not justify the result presented in
Fig.~2 of [1] where only few holes are involved, and ii) it should
imply an unjustified increase of $L_{\rm SPP}$ by 2 or 3 orders of
magnitude necessary for the SPP to travel through the hole arrays of
Figs.~3,4 of [1]. Outside the array, the SPPs still follow the
dispersion relation of the unperturbed film and therefore have to be
excluded from the image formation, due to their low propagation
length.

\newpage
\begin{figure}[h]
\includegraphics[width=8cm]{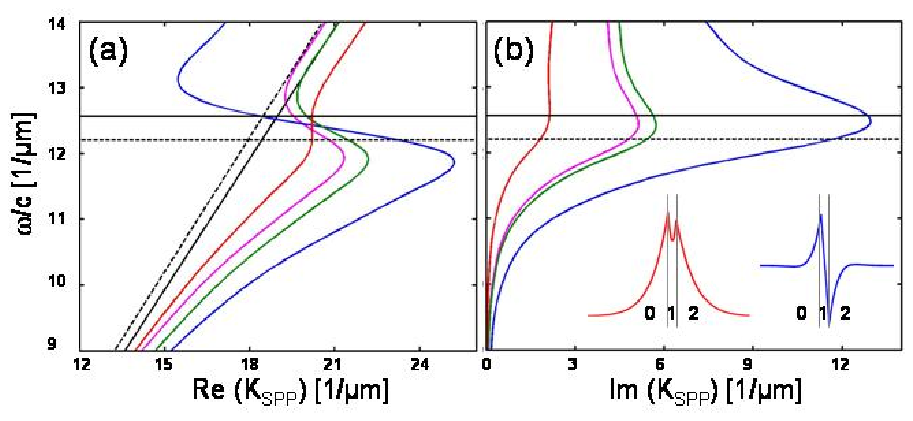}
\end{figure}

figure 1
\newpage

Caption for figure1:

 (color online) Dispersion relation for bound
SPPs propagating on a 50 nm thick gold film (medium 1)  surrounded
by glass (medium 2) and glycerin (medium 0). (a) shows the
dispersion for the real part of the SPP wavevectors while (b) shows
the imaginary part. The red and blues curves correspond to the two
bound modes (see inset in (b) for the tangential magnetic field
profiles). The green and magenta curves show the dispersion for SPPs
bound at single interfaces gold/glass and gold/glycerin,
respectively. The horizontal black lines show the range of frequency
used in [1,2,3]. The light lines for glass (continuous) and glycerin
(dashed) are included in (a).
\end{document}